\documentclass[preprint2]{aastex}

\newcommand{\vdag}{(v)^\dagger}

\slugcomment{accepted for the ApJ: 2004 Jul 14; fix typos/spellings 16jul}

\shorttitle{The Jet of 3C120}
\shortauthors{Harris et al.}

\begin{document}


\title{The X-ray Jet of 3C 120: Evidence for a Non-standard
  Synchrotron Spectrum}


\author{D. E. Harris, A. E. Mossman}

\affil{Smithsonian Astrophysical Observatory, Harvard- Smithsonian
Center for Astrophysics 60 Garden St., Cambridge, MA 02138
\email{harris@cfa.harvard.edu} \email{amossman@cfa.harvard.edu}}

\and

\author{R. C. Walker\footnote{The
National Radio Astronomy Observatory is operated by Associated
Universities, Inc., under contract with the National Science
Foundation.}}

\affil{National Radio Astronomy Observatory, Box 0, Socorro,
NM 87801\email{cwalker@nrao.edu}}





\begin{abstract}
We report on archival data from the Chandra X-ray Observatory for the
radio jet of 3C120.  We consider the emission process responsible for
the X-rays from 4 knots using spectra constructed from radio, optical,
and X-ray intensities.  While a simple synchrotron model is adequate
for three of the knots, the fourth ('k25'), which was previously
detected by ROSAT and is now well resolved with Chandra, still
represents a problem for the conventional emission processes.  If, as
we argue, the flat X-ray spectra from two parts of k25 are synchrotron
emission, then it appears that either the emission comes from an
electron distribution spectrally distinct from that responsible for
the radio emission, or at the highest electron energies, there is a
significant deviation from the power law describing the electron
distribution at lower energies.
\end{abstract}




\section{Introduction\label{sec:intro}}

3C120 is a nearby (z=0.033) radio galaxy which has variously been
classified as a Seyfert 1, a broad line radio galaxy, and a late type
spiral.  The optical appearance is complex, with multiple dust lanes
and two suggestive spiral arms which become radial features at a
projected distance of 6.3 kpc from the core (fig.~\ref{fig:k4hst}).
The radio morphology and luminosity are more similar to the FRI class
radio galaxy (Fanaroff \& Riley 1974) than to the usual Seyfert.  A
one sided pc scale jet has been extensively studied with long baseline
interferometry and superluminal components have been identified
(apparent speeds between 3 and 6c, Walker et al. 2001).  The kpc scale
jet is shown in fig.~\ref{fig:lband} and there is additional complex
emission at larger scales (beyond the borders of fig.~\ref{fig:lband},
Walker, Benson, \& Unwin 1987).

At X-ray wavelengths, the nucleus is a strong and variable source
(e.g. Halpern 1985), but it was the detection with ROSAT of a rather
inconspicuous radio knot 25$''$ from the core (16 kpc, projected),
which presented new problems for the conventional models of X-ray
emission from radio jets (Harris et al. 1999; HHSSV hereafter).
Therefore we obtained new radio data at higher frequencies to see if
this knot was peculiar or unique either in shock structure (as
manifest by gradients in radio surface brightness) or in radio
spectra.  Now that Chandra data are in the archive, we have
additionally studied the inner jet as well as being able to resolve
the 25$''$ knot, so that we separate X-ray components by their
morphology, and then test the various emission models.

The following sections describe the observations, data reduction, and
photometry (\ref{sec:obs}), the radio results (\ref{sec:results}), the
X-ray emission processes (\ref{sec:emis}), and an evaluation of each
detected X-ray feature (\ref{sec:eval}).  We label the knots in the
jet which are of primary interest, by their distance from the core in
arcseconds (e.g. 'k4', 'k25').  We take the distance to 3C 120 to be
140 Mpc (H$_{\rm{o}}$=71~km~s$^{-1}$~Mpc$^{-1}$; Spergel et al. 2003)
so that 1" corresponds to 0.64 kpc.  We use the standard definition of
spectral index, $\alpha$: flux density, S$_{\nu}\propto\nu^{-\alpha}$.

\section{The observations and reduction procedures\label{sec:obs}}

\subsection{The Radio Data}\label{sec:radio}

Details of the observations used in this paper are presented in Table~\ref{tab:vla}. 
A comprehensive set of VLA and VLBI images between 1.4 and 15 GHz was
published by Walker, Benson, and Unwin (1987).  Here we use the 1983
October 2 VLA A-configuration image from that paper as the source of
our 1.4 GHz flux densities.  An extremely high dynamic range image of
3C~120 at 5 GHz was produced by Walker (1997) as a byproduct of an
effort to measure the proper motion of the 4\arcsec\ knot.  We use
that image, which is actually the weighted sum of 3 images widely
separated in time (no variability in the jet was detected), to measure
our 5 GHz flux densities.

To obtain good images at appropriate resolution at higher frequencies,
we observed 3C120 at 14.9 GHz using the VLA B-configuration (second
largest) and at 22.2 and 43.3 GHz using the VLA D-configuration
(smallest).  Both 22.2 and 43.3 GHz were observed with subarrays
because 43.3 GHz was only available on 12 antennas.  The 22.2 GHz
image has inadequate resolution for use in this project and will not
be presented.  Reference pointing was used to keep 3C120 positioned
properly in the VLA primary beams at 43 GHz.  Because we were
interested in the knot offset by 25\arcsec\ from the core, and the VLA
primary antenna beam is only about an arcminute FWHM at 43 GHz, we
offset the pointing position by 12\arcsec\ in the direction of the
knot at that frequency.  For all images, baseline based corrections
for closure errors were made using scans on the bright, compact source
3C~84.  The procedure is described in the above references.

All of the VLA images were calibrated following the normal procedures
as described in the VLA Calibrator Manual.  For all observations, the
VLA continuum system was used with 50 MHz in each of two polarizations
in each of two channels (200 MHz total).  All data reduction was done
using the NRAO Astronomical Imaging Processing System (AIPS).  A
primary beam correction was made for the 15 and 43 GHz images.  At 43
GHz, that correction took into account the offset pointing position.

The measured flux densities for each of the features of interest are
presented in Table~\ref{tab:fluxden} along with the optical and X-ray
values.  We estimate that the errors in our measured flux densities of
strong features are 10, 5, 6, and 10\% at 1.4, 5, 15, and 43 GHz,
respectively.  To this we add the image noise times the square root of
the number of beam areas over which a given flux density is measured.
The radio flux densities were measured in exactly the same boxes from
each of the radio images although these rectangles, aligned with RA
and DEC, are not identical to those used for optical and X-ray
photometry.  This is not thought to be a significant source of error
since we ensured that the same features were being measured on the
basis of radio overlays.

The resolutions of the images fall between 0.36\arcsec\ and about
2\arcsec.  Normally it is a poor idea to use interferometric images
with unmatched resolutions for the measurement of flux densities from
which spectra are to be derived.  But in this case, there is no large
scale structure in the regions of interest that is likely to show up
in some images and not in others.  The high resolution images are
of high quality so all the flux density that would be seen at lower
resolution should be present.  As a result, we do not consider the
mismatched resolutions to be a significant limitation.  In fact, the
normal effect of such a mismatch would be to have too little flux
density measured in the higher resolution images.  An examination of
the spectra shows that, if anything, the opposite situation exists
here.

\subsection{The optical data\label{sec:opt}}

We retrieved an HST image of 3C120 from the WFPC2 Associations
website.  The selected file was taken from the F675W filter with a
pivot wavelength of 6717 \AA. We registered the data so as to align
the optical and radio peak intensities (0.72$''$ shift).  For the HST
data, the peak position was found by the intersection of the
diffraction spikes.  After constructing a flux map
(fig.~\ref{fig:k4hst}) we measured the features of interest: k4, k7,
s2, s3, and k25inner.  The other parts of k25 were off the edge of the
image, as was k80.

Uncertainties were estimated by using the same regions on the events
image.  For both sets of measurements we used 'funtools'\footnote{
(http://hea-www.harvard.edu/RD/funtools/)}.  For k4, which lies on a
relatively bright optical spiral like feature, we used a circular
aperture significantly smaller than that used in the radio and X-ray
bands (r=0.273$''$), chosen so as to include all of the obvious optical
emission of k4. For the background, we chose the same size circle,
positioned just to the North of k4 where the brightness was judged to
be about the same as that surrounding k4.  Note however, that this
subjective choice means that the actual uncertainty in the flux of k4
is larger than the statistical error quoted in Table~\ref{tab:fluxden}.

For the remaining features we used the same regions as defined for the
X-ray photometry, except that we used a different background region for
k25inner since the original background region was off the image.  The
resulting flux densities are given in Table~\ref{tab:fluxden}.

For k25in which is close to the edge of the image, we found -58 counts
in the source region (the background was oversubtracted), but +40 and
+58 counts in two equally sized rectangles above and below the source.
Thus we take $\sigma$ to be about the measured flux
value: 0.34$\times10^{-29}$ (cgs).  For an upper limit, we take 3
$\sigma$ or 1 $\mu$Jy.

\subsection{The X-ray Data}\label{sec:xray}

The archival Chandra data analyzed consist of an ACIS-S imaging
observation (obsid 1613) and the zeroth order of an HETG observation
(obsid 3015).  The detectors are described in the CHANDRA Proposer's
Observatory Guide\footnote{Rev.6.0, December
2003. http://cxc.harvard.edu} and the parameters of the observations
are given in Table~\ref{tab:xlog}.

The imaging data were badly saturated at the core position, whereas the zero
order image of the HETG data did not suffer from this problem: the so
called ``level 1'' event file had a peak intensity pixel of 4940
counts (0.086 c/s, or 0.28 c/frame-time).  Thus there is some pileup,
but no saturation.  The imaging observation also was done with a roll
angle that made the readout streak fall on the inner jet (out to k4
and k7), but k25 was not affected.  For these reasons, most of our
results are based on the zero order image.

Our basic data reduction follows the procedures we used for M87 data
(Harris et al. 2003).  After checking the background for particle
flares (our primary dataset was clean, and we decided not to exorcise
the high countrate segment of 1613), we removed pixel randomization
and registered the event file so that the x-ray and radio cores were
aligned.  This process for 1613 was problematic since the location of
the X-ray core had to be estimated from the readout streak in one
dimension, and for the orthogonal direction, an alignment of a circle
with the wings of the central count distribution.

We then created 1024 arrays binned into 1/10 native ACIS pixel for
each of four energy bands; constructed monochromatic exposure maps for
an energy near the center of each band; and produced 'flux maps' with
units of ergs cm$^{-2}$~s$^{-1}$ by dividing the data by the exposure
map and multiplying by the nominal h$\nu$ of the band.  The exposure
maps were generated using the time dependent effective area files
which deal with the buildup of contamination on the ACIS detector.
An example of a flux map is shown in fig.~\ref{fig:regions} with the 
regions overlayed.

Flux measurements consisted of choosing appropriate regions for jet
features and corresponding backgrounds (see table~\ref{tab:regions}
and fig.\ref{fig:regions}); measuring the flux; and correcting for the
mean energy (via the same source region, but operating on the event
file).  Since the flux maps were constructed with a multiplicative
factor of h$\nu_{nominal}$, the correction factor of $<E>/E_{nominal}$
for each measurement is required to recover the actual flux within the
band. In Table~\ref{tab:xflux} we give the counts and flux for each
jet component of interest.

Conversion of flux to flux density was done for an assumed power
law with $\alpha$=1.  Although this procedure is not strictly correct,
the error introduced by using an incorrect value of $\alpha$ should be
of second order since the bands were only a few keV wide.  In general,
we did not have enough counts to obtain an accurate value of
$\alpha_x$, and for the brighter features such as k4, $\alpha_x$ is
not far from 1.

Quoted uncertainties in fluxes and flux densities are purely
statistical.  They were derived by using the source and background
regions on count maps (i.e. before dividing by the exposure maps).
The fractional error in each flux was then carried forward to the flux
densities.

Since most features do not have sufficient counts to provide an
accurate estimate of the spectral parameters using the normal fitting
procedures in the X-ray band, we use our measurements of flux
densities (Table~\ref{tab:fluxden} ) to construct spectra (see
sec.~\ref{sec:eval} for examples).  Except for the uncertainty in the
absorption correction for the softest band, we believe that our
technique of deriving spectral parameters from the flux maps is
reliable.  For the case of k4 there are sufficient counts to allow the
usual X-ray method of spectral fitting.  We did that with Ciao/Sherpa
and find agreement between the two methods, both in values of the
amplitude and $\alpha_x$.

\section{Results from the new radio data\label{sec:results}}

We planned and obtained our new VLA radio data on the basis of the
ROSAT observations of 3C 120.  At that time a persistent question was
``Why are a few knots and hotspots in radio jets detected with current
X-ray sensitivities while the vast majority are not?''  We now know
that this question has been mostly answered because of the large
number of jet detections with
Chandra\footnote{\url{http://hea-www.harvard.edu/XJET/}}: with
sufficient sensitivity and angular resolution, X-ray detections of
radio jets are quite common and no longer drive us to search for
'special conditions'.  However, it would still be useful to identify
any peculiarities in shock regions containing electrons which produce
significant X-ray emission.  For k25 in particular, the difficulty of
finding a reasonable emission mechanism for the X-rays led HHSSV to
suggest the possibility of a flatter spectrum (than deduced from the
radio data) population of relativistic electrons extending up to
Lorentz energies of $\gamma \approx 10^7$.

For these reasons, we undertook high frequency VLA observations to
better determine the radio spectrum and to evaluate the behavior of
the sharp gradients in radio surface brightness at two locations in
k25.  Both of these parameters may be observational signatures of
physical attributes of shocks.

\subsection{Surface Brightness gradients}

We used the AIPS tool NINER to measure the first derivative of the
radio brightness.  In order to measure comparative gradients at
different frequencies, maps were convolved to the largest beamsize.
However, when comparing different features on the same map, this
convolution was not necessary.

For all comparisons we computed the normalized gradient by dividing
the actual value of Jy/beam/pix by the intensity of the peak brightness of
the feature being measured.  When comparing maps with different pixel
size, we also divided by the factor arcsec/pixel.  In addition to
measuring the gradient for the outer edge of k25, we measured
gradients for the core (which serves to define the value for an
unresolved source), k4, and the leading edge of
k25 (which was undetected in our ROSAT data but is clearly
present in the Chandra images).

\subsection{Flux densities for spectra}

To compare the spectra of the inner and outer shocks of k25, we
measured flux densities on the Walker et al. (1987) maps constructed
to have the same beamsize of 1.25$"$ FWHM (1.4, 5, and 15 GHz) and
then repeated the measurements on maps smoothed to match the
resolution of the 43 GHz map (2.2$''\times1.6''$) in order to include
the 43 GHz flux density.  We defined a rectangle of 3$''\times2.4''$
(RA$\times$DEC) for the internal (southern) shock and 2.4$''\times3''$
for the outer (western) shock.  Although these integration areas are
small, we corrected for background level by running IMEAN on a nearby
region 15.6$''\times13.2''$.  These corrections were $<$ 6\% in all
cases.  The results are given in Table~\ref{tab:fluxden}.  The
resulting spectral indices are $\alpha_{out}$~=~0.66$\pm$0.03 and
$\alpha_{in}$~=~0.74$\pm$0.06.

\subsection{Evaluation of Radio Results}

There is no clear evidence for a significant difference in the
spectral indices of the two parts of k25.  The values are very close
to each other and there are no data that demand a spectral break
between 1.4 and 43 GHz.  The outer shock structure is brighter than
the inner feature at all frequencies we have measured.

Our attempts to measure the magnitudes of the gradients in the surface
brightness have been less than rewarding.  To make progress with this
approach, it would be necessary to have a resolution and sensitivity
at least equal to those of fig.~\ref{fig:regions}.  Our only
indicative conclusion is that the ratio of the normalized gradients of
the outer to the inner shock structures is greater than one at and
below 5 GHz, and less than one above 10 GHz.  However, the beamsizes
employed even for this comparison are judged to be too large for any
reasonable accuracy.

We conclude that there is very little or no difference between the two
shocks (except for the total flux density) and neither is discernibly
exotic.  Given the ever increasing number of X-ray detections of radio
jets, we take the absence of evidence for peculiarity in these
putative shock features to be consistent with the idea that X-ray
emission is relatively common for radio jets with a reasonable
velocity towards us (i.e. two-sided X-ray jets are rare; we cannot
cite even a single convincing example).

\section{X-ray Emission Processes\label{sec:emis}}

In this section we discuss general aspects of the possible emission
processes and for each jet feature tabulate the key parameters for
each process.  In section~\ref{sec:eval} we compare the processes for
each feature in turn.  

There are two basic problems: how to estimate the emitting volume
(which influences the calculated value of the equipartition
magnetic field strength) and what constraints might there be on the
Doppler beaming factor, $\delta$ (which governs the IC calculations).
In general, we have elected to use the largest reasonable volume,
with the X-ray and highest resolution radio maps as a guide.
Table~\ref{tab:regions} provides the dimensions and shapes of our
assumptions.  In some or possibly all cases the actual volumes may be
smaller than that assumed, and that would lead to larger equipartition
magnetic field strengths for synchrotron models or higher electron
densities for thermal bremsstrahlung models.

To accommodate relativistic beaming from bulk jet fluid velocities
($\Gamma=(1-\beta^2)^{-1/2}$; $\beta$=v/c), we will derive
synchrotron parameters for a few representative values of
$\delta=1/\Gamma(1-\beta~cos\theta)$ ($\theta$ is the angle
between the jet direction and the line of sight).  For thermal models
we assume there is no beaming and for IC/CMB models we solve for the
characteristic value of $\theta$ for which $\Gamma=\delta$.

\subsection{Synchrotron models}\label{sec:sync}

We found no evidence that the outer shock in k25 is discernibly exotic
in the radio
(section~\ref{sec:results}), but from the X-ray morphology of k25, it
is evident that the X-ray and radio emissions are co-spatial.
This demonstrates that the suggestion of HHSSV,
that the X-ray emission might originate in a much smaller volume than
the radio emission, is not supported by the new data.

The synchrotron parameters listed in Table~\ref{tab:sync} are typical
for FRI jets.  With beaming, the intrinsic power of the source drops.
Since we posit no change in emitting volume, the field drops as well.
The halflife for electrons responsible for 2 keV emission ($\tau_{\frac{1}{2}}$) increases
as the field drops, but eventually the IC losses dominate and so
$\tau_{\frac{1}{2}}$ decreases.  Since we have chosen to use the largest
allowable emitting volume in each case, the actual value of the magnetic
field strength may be larger than indicated, and this would
decrease $\tau_{\frac{1}{2}}$.

The classical objection to the synchrotron model arises for those
features which have an obvious inflection in their spectra to
accommodate a flatter X-ray spectrum than would be anticipated by the
segment connecting to optical or radio data.  Two possible escapes
from this objection are: (a) the existence of a distinct spectral
component which is characterized by a flatter electron distribution so
that its emission is below detection limits at frequencies below the
X-ray band (as suggested for k25 by HHSSV), and (b) a scheme that
leads to a 'pileup' of electrons near the cutoff energy (e.g. Dermer
\& Atoyan 2002).

\subsection{Beaming models}\label{sec:ic}

Inverse Compton emission from scattering on CMB photons by jet
features with relativistic beaming has been suggested as a method to
increase the energy produced via the IC channel relative to that in
the synchrotron channel (Tavechhio et al. 2000; Celotti et al. 2001).
Harris and Krawczynski (2002) calculated beaming parameters
implied from the 3C~120 ROSAT data (HHSSV) if the jet components are in equipartition
(energy density, u(particles) $\approx$ u(B)).  With our revised values for the
spectral indices and flux densities for several other features, we
have recalculated the beaming parameters (Table~\ref{tab:ic}).

The beaming parameters given in Table~\ref{tab:ic} are not entirely
unreasonable if the $\Gamma$ of the jet does not decrease
significantly out to k25.  However, as noted in the table, for several
features the spectral index observed in the X-ray band disagrees with
that expected from the radio spectrum.  Additionally, the small angles
of the jet to the line of sight (of order 5$^{\circ}$) for most of the
knots, leads one to infer a large projection effect so that the
physical size of the jet becomes rather long for an FRI.  Furthermore, the
beaming parameters required ($\Gamma, \delta, \theta$) are more
extreme than those generally applied to the pc scale jet (Gomez et
al. 2000; Walker et al. 1987).  

A more general problem for the IC/CMB model is why the jet is not
continuous in the X-rays once the emission 'starts' with the
generation of particles at a shock.  Since we know the jet fluid
continues downstream from a knot, the only way to quench the IC
emission from the low energy electrons responsible for the X-rays, is
to deflect the jet so that it is no longer pointing towards us, or to
drastically reduce its velocity so that $\Gamma$ drops back close to
one (in which case it would have to increase again at the next knot).
For the synchrotron model, X-ray knots end quickly because of the
large E$^2$ losses to the extremely high energy electrons required,
but the very low energy electrons responsible for X-rays in the IC/CMB
model have a very long half-life and thus should propagate
undiminished for sizable segments of the jet.  These and other
difficulties of the IC/CMB models are discussed by Atoyan \& Dermer
(2004) and Stawarz et al. (2004).  We believe the weight of the
evidence argues against the IC/CMB model for the X-ray emission of the
3C120 jet.

\subsection{Thermal Bremsstrahlung}\label{sec:thermal}

Given the flat spectra found for a few features, we consider the
possibility that the X-ray emission comes from hot gas (c.f. the
interpretation for NGC 6251 by Mack, Kerp, \& Klein 1997).  For rough
estimates of the parameters, we calculate the (uniform) density
required to produce the observed X-ray flux.  The results are given in
Table~\ref{tab:thermal}.

The standard arguments against the thermal bremsstrahlung emission process include:

\begin{itemize}

\item{ if the X-ray emission occupies the same volume as the radio
emission, there would be a departure from the $\lambda^2$ law for the
radio polarization position angle;}

\item{if the thermal region surrounds the emitting region, the predicted
large rotation measures (RM's) are not seen; and}

\item{the thermal pressure of the emitting region generally exceeds the estimate
of the ambient gas pressure.}

\end{itemize}

At least for the inner jet, the observed RM's are small (0 to -10
radians~m$^{-2}$, Walker et al. 1987), so that the large RM's of
Table~\ref{tab:thermal} would appear to preclude thermal X-ray
emission.  Reliable RM's are not available for k25, but from
polarization position angles at two frequencies for the brighter parts
of k25, it seems likely that RM$\leq$100~rad~~m$^{-2}$.  No
polarization data are available for k25new because of the very low
radio brightness of this region so part, or conceivably all, of the
X-ray emission from k25new could be from hot gas.

\section{Evaluation of Features}\label{sec:eval}

\subsection{The inner jet}

As noted earlier, the X-ray image from the archival ACIS data without
the grating is heavily saturated at the core position so that it is
impossible to study the innermost region of the jet.  Even the HETG
zero order image has some degree of pileup, which means that the peak
of the PSF is depressed relative to the wings.  In spite of these
difficulties, we constructed a PSF with ChaRT (the Chandra ray tracer)
and subtracted it from the image.  From a series of trial runs using
different normalizations, we are convinced that there is substantial
excess emission to the SW of the nucleus (and here we refer to the
position of the radio nucleus since we registered all our files so as
to align the radio and X-ray peaks).  This emission lies between
0.5$^{\prime\prime}$ and 0.9$^{\prime\prime}$ from the nucleus at a PA
of $\approx~225^{\circ}$; i.e. projected on the southern edge of the
radio jet.  An example of one of the residual maps is shown in
fig.~\ref{fig:inner}.  Given the mismatch between the generated PSF
and the data, we will not pursue the attributes of this feature here.

The X-ray emission from k4 is well resolved from the core region and
aligns well with the corresponding optical and radio emissions
(figs.~\ref{fig:k4hst} and~\ref{fig:inner}).  The values of the X-ray
spectral index, $\alpha_x$, from spectral fitting (0.95$\pm$0.45) and
from our 4 band fluxes (0.90$\pm$0.20) agree within the errors and are
also consistent with the radio to X-ray index, $\alpha_{rx}$=0.90.
Furthermore, the optical flux density falls on the power law defined
by the radio and X-ray data.  Thus we favor a synchrotron explanation
(perhaps with modest beaming, e.g. $\delta$=3) for the X-ray emission.
If the X-rays were to be beamed IC/CMB radiation, the apparently
straight power law fit from radio and optical to X-rays would be a
coincidence and $\delta$ would have to be $>$10 (Table~\ref{tab:ic}).
We note that while it appears that k4 lies in a spiral arm feature of
the galaxy, we consider it highly probable that this is a coincidence.
At the distance of 3C~120, 4$''$ corresponds to 2.5 kpc, a reasonable
distance between the nucleus and a feature defined by stars.  However,
from VLBI studies of the superluminal radio jet, the angle between the
jet and our line of sight is of order 5$^{\circ}$, so k4 would be
$\approx$~29~kpc from the nucleus, and thus starlight would not be a
viable source of photons for IC scattering.

\begin{figure}
\plottwo{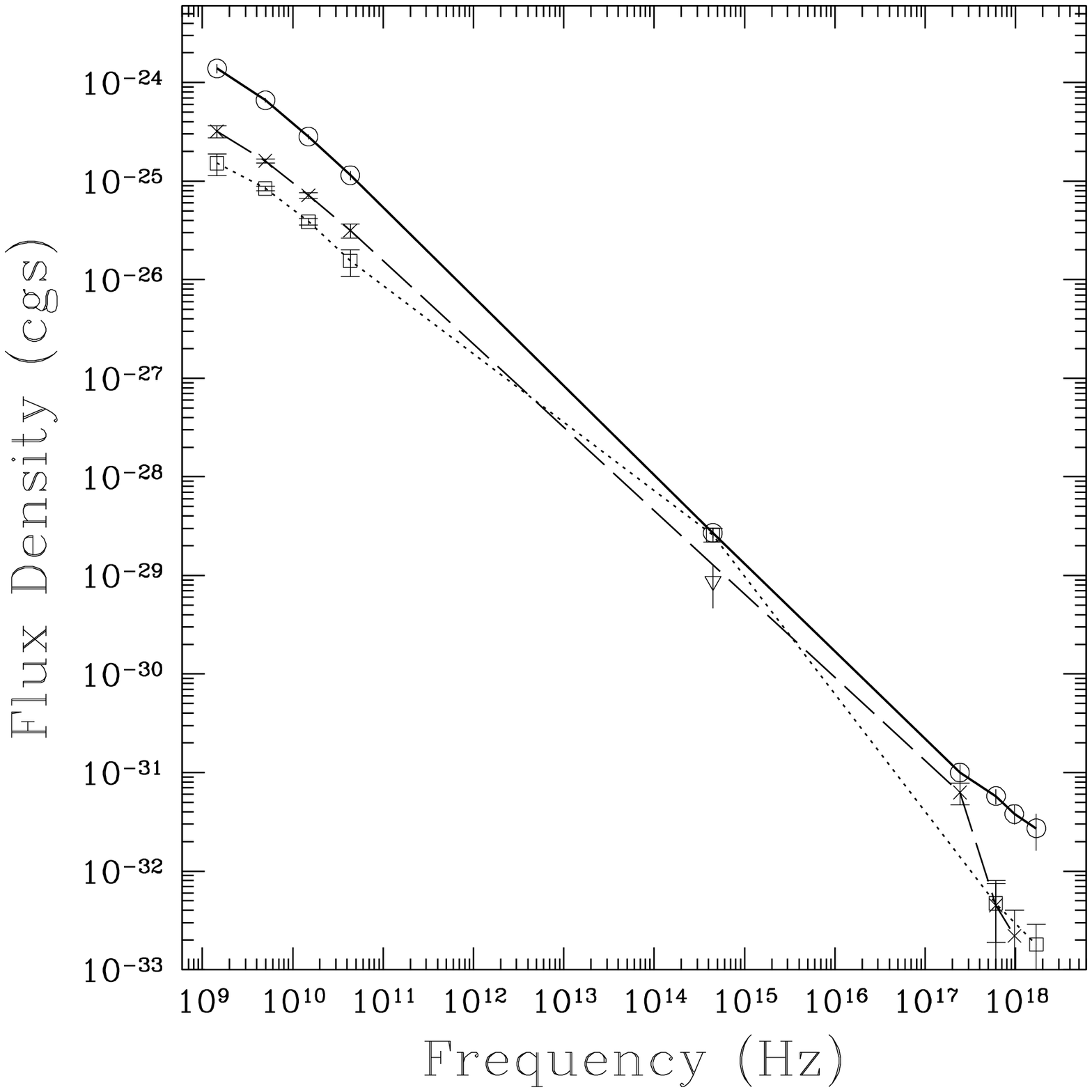}{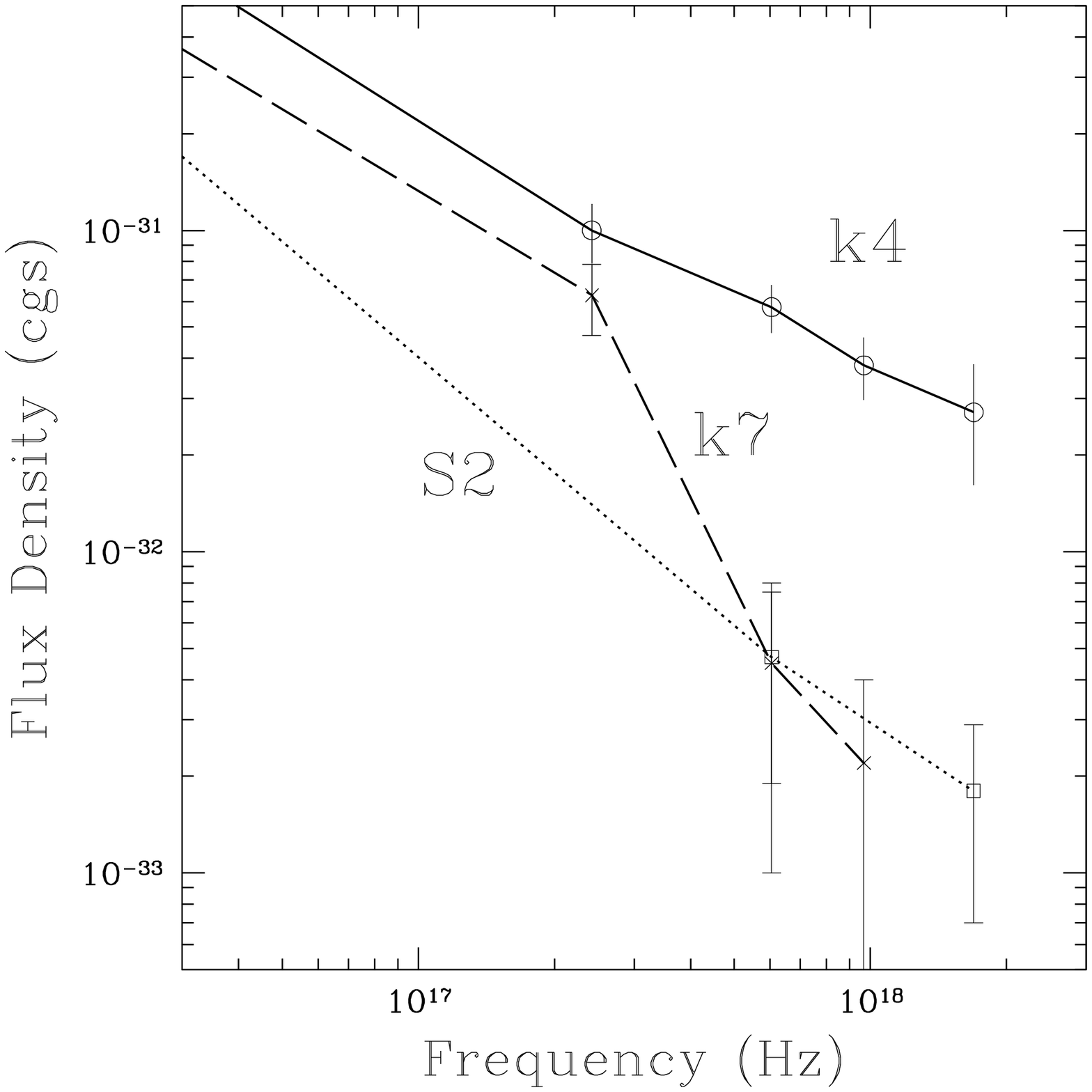}
\caption{Spectra for components of the inner jet.  The solid line
(circles) is for k4; the dashed line (crosses), k7; and the dotted line
(squares), s2.  The down triangle is a 2 $\sigma$ (optical) measurement for k7.
The optical values for k4 and s2 are so close to each other that their
symbols overlap.  The left panel shows the radio to X-ray spectrum;
the right panel shows the X-ray data.
\label{fig:specin}}
\end{figure}

The next knot, k7, has only 30 net counts so no spectral analysis was
attempted. However, as shown in fig.~\ref{fig:specin}, there is a
significant difference between $\alpha_{rx}$=0.86 and
$\alpha_x$=2.4$\pm$0.6, as derived from flux densities for 3 of our
energy bands.  This sort of spectrum is consistent with a power law
distribution of relativistic electrons with a high energy cutoff
(i.e. suffering E$^2$ energy losses); so again, we favor the
synchrotron model.  The 2$\sigma$ optical detection falls on the power
law connecting the radio and X-ray data.

The next section of the jet ("s2" in fig.~\ref{fig:regions}) is very
weak in the X-rays and the following section, "s3", was not detected.
However, the measurements are reported in
Tables~\ref{tab:fluxden}~and~\ref{tab:xflux}.

\subsection{Knots k25 and k80}

The X-ray emission of k25 continues to be a problem insofar as the
standard emission models are concerned (HHSSV).  The new data make
matters worse since we now find extended X-ray emission in a region of
very low radio brightness.  We call this region 'new', and use the
terms 'inner' and 'outer' for the two relatively bright radio edges
which show large gradients in surface brightness: 'inner' refers to
the upstream edge and 'outer' refers to the western edge.  In
fig.~\ref{fig:k25marbles} we show the events comprising k25.

\begin{figure}
\plottwo{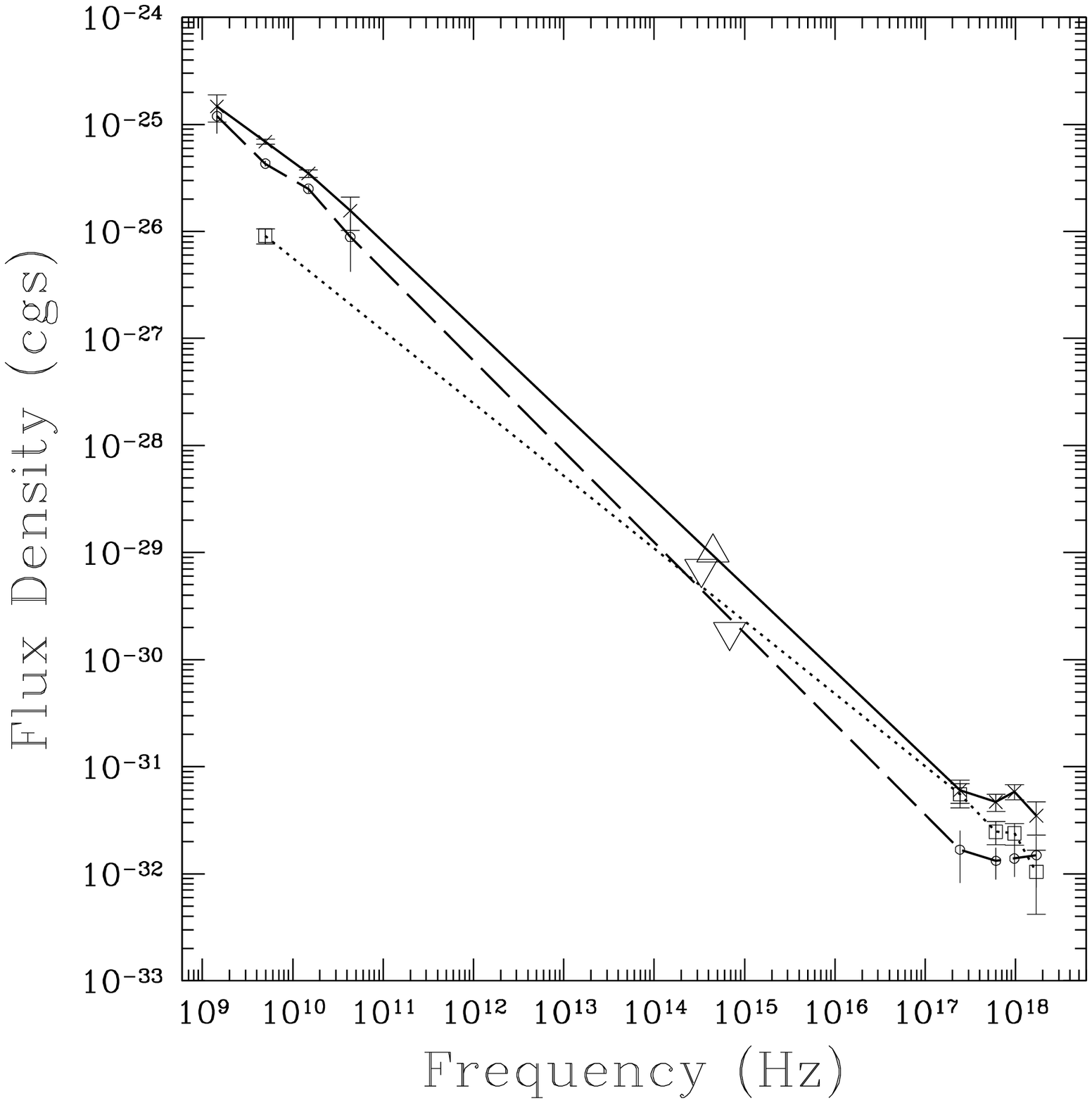}{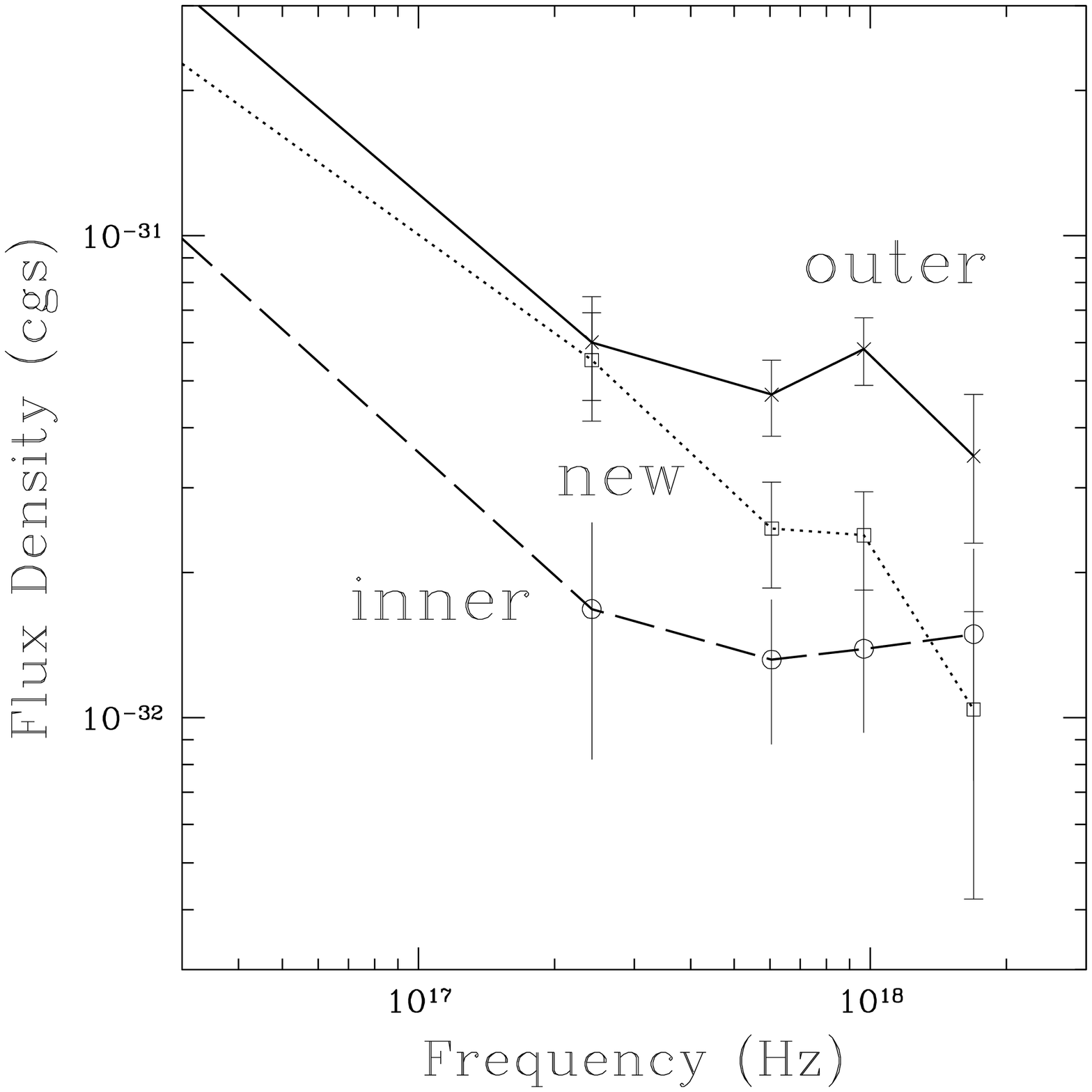}
\caption{The spectra for the components of k25.  The left panel shows
  the radio to X-ray spectrum; the right panel shows the X-ray data.
  The strongest feature is the outer region, crosses and solid line;
  the inner region is marked by circles and a dashed line; and the
  'new' region is denoted by squares and a dotted line. The
  up-triangle is an upper limit for k25in from HST, and the two
  down-triangles are upper limits reported by HHSSV for the total k25
  optical emission. \label{fig:speck25}}
\end{figure}

The only optical data we have is the upper limit to k25in, and this
flux density lies well above the power law connecting the radio and
X-ray flux densities.  Note however that the optical upper limits
quoted in HHSSV for the area then known to be associated with X-ray
emission fall below the spectrum of the total emission, so the
suspicion remains that it is not feasible to argue for a single power
law from the radio to the X-ray.  As shown in fig.~\ref{fig:speck25},
both the inner and outer regions have quite flat X-ray spectra,
whereas $\alpha_x$ for the 'new' region is consistent, within the
errors, with $\alpha_{rx}$.

A flat power law ($\alpha_x$ close to zero) is normally not
considered feasible as a synchrotron spectrum associated with
standard shock acceleration expectations, but what are the 
alternatives?

Aside from the large beaming factor required for the IC/CMB model
($\delta>$10, Table~\ref{tab:ic}), one would need to explain why the
emission dies away moving downstream since the very low energy
electrons responsible for IC/CMB should have very long halflives.  The
major unknowns for the IC/CMB model are the validity of extrapolating
the electron spectrum down to the required low values of $\gamma$, and
the actual value of the bulk relativistic velocity, i.e. $\Gamma$.

As shown in Table~\ref{tab:thermal}, thermal emission models encounter
a number of problems and thus we do not believe that thermal bremsstrahlung 
is a viable alternative to synchrotron emission for the
brighter regions, k25in and k25out.

For the synchrotron model, aside from the requirement for a separate
spectral component with a flat spectrum, we face the difficulty of
propagating high energy electrons from the acceleration sites
(i.e. the inner and outer large-gradient radio features) to fill the
enclosed volume ('new').  The required distance is of order 3$''$ (1.9
kpc in projection); a light travel time of some 6000 years.  This may
be compared with typical lifetimes of the highest energy electrons
against E$^2$ losses (with equipartition magnetic field strengths) of
$\approx$~200 years (Table~\ref{tab:sync}). If the X-ray emission from
k25new is thermal (the anonymous referee suggested shocked gas), the
travel time problem for high energy electrons would be moot.

An alternative explanation for the presence of high energy electrons
in the 'new' region would be distributed acceleration (i.e. not
directly associated with obvious shocks), as might be the case for
reconnection scenarios (Birk \& Lesch 2000, Birk, Crusius-W\"{a}tzel,
\& Lesch 2001), or acceleration associated with
turbulence along the jet edge (Stawarz et al. 2004, and references therein).

Knot k80 (fig.~\ref{fig:lband}) is of low radio brightness, but we
find associated X-ray emission.  The morphological correspondence
between radio and X-rays is not close as in the other 3 knots.  In
fig.~\ref{fig:k80} we show the relevant maps.  Insofar as the detected
structures are concerned, the situation of the X-rays being slightly
displaced upstream of the bulk of the radio emission is very similar
to that in the outer knot in the jet of PKS1127-145 (Siemiginowska et
al. 2002).  The three values of S$_x$ in Table~\ref{tab:fluxden} are
consistent with $\alpha_{rx}$=0.8.  As for the inner knots, the
results of section~\ref{sec:emis} leave little doubt that the
x-rays come from synchrotron emission.

\section{Conclusions}

The X-ray emission from k25
remains a puzzle.  The only ad-hoc models which we can suggest involve
distinct spectral populations of relativistic electrons.  For
synchrotron emission, this would be a flat spectrum component
extending to electron Lorentz factors of $\gamma=10^7$ to $10^8$ (see table 5
of HHSSV).  Note however that ``a distinct spectral population'' need
not indicate a separate shock region for its genesis.  Dermer \&
Atoyan (2002) have described E$^2$ loss conditions which can produce
spectral hardening at high energies near the cutoff, and Niemiec \&
Ostrowski (2004) have produced numerical simulations for what they
call 'realistic' magnetic field structures in relativistic shocks that
can provide both relatively flat particle distributions and even
flatter high energy tails for energies below the cutoff.  For IC/CMB
with beaming, the parameters derived above could be relaxed if an
additional, steep spectrum distribution of electrons existed below
$\gamma=2000$ (see section 5.2 of Harris and Krawczynski 2002).
A 60ks observation with Chandra has been approved for AO6.  When these
data are available we should obtain more accurate X-ray spectral
parameters.

Although one of us (DEH) has been a long and ardent supporter of the
classical view: ``If the spectrum is not concave downward, it is not
synchrotron emission'', it is now our view that there is strong
evidence for marked deviations from this scenario.  It should be
remembered at this point, that there are many other knots in jets for
which $\alpha_x~<~\alpha_{ox}$, (e.g. NGC6251, Sambruna et al. 2004),
and we believe it would be more fruitful to re-interpret these
occurrences rather than to cite them as proof that synchrotron models
are impossible.

\acknowledgments

The work at the CfA has been partially supported by NASA contract
NAS8-39073 and NASA grant GO3-4124A.  We thank the creators of the
WFPC2 Associations\\ (http://archive.stsci.edu/hst/wfpc2) for
providing public access to useful data.  Tahir Yaqoob generously
shared the zero order image of his proprietary data (we joined him in
a Chandra AO-4 proposal for an additional observation of 3C 120, but
were unsuccessful).  We thank C. Cheung for his careful reading of the
manuscript and his resulting suggestions.  We also thank the anonymous
referee for his useful comments.

\clearpage

\section{References}

\noindent
Atoyan, A. \& Dermer, C. D. 2004, (submitted to ApJ)

\noindent
Birk, G. T. \& Lesch, H. 2000 ApJ 530, L77

\noindent
Birk, G. T., Crusius-W\"{a}tzel, A. R., \& Lesch, H. 2001 ApJ 559,
96

\noindent
Celotti, A., Ghisellini, G., \& Chiaberge, M. 2001 MNRAS 321, L1

\noindent
Dermer, C. D.\& Atoyan, A. M. 2002 ApJ 586, L81

\noindent
Gomez, J. L., Marscher, A. P., Alberdi, A., Jorstad, S. G., \&
Garcia-Mir\'{o}, C. 2000, Science 289, 2317

\noindent
Halpern, J. P. 1985 ApJ 290, 130.

\noindent
Harris,  D. E. and Krawczynski, H.
2002, ApJ 565, 244

\noindent
Harris, D.E., Hjorth, J., Sadun, A.C., Silverman, J.D. \& Vestergaard, M.
1999 ApJ 518, 213 (HHSSV)

\noindent
Harris, D. E., Biretta, J. A., Junor, W., Perlman, E. S., Sparks,
W. B., and Wilson, A. S. 2003 ApJ 586, L41.

\noindent
Hjorth, J., Vestergaard, M., S$\o$rensen, A. N., \& Grundahl, F. 1995
ApJ 452, L17

\noindent
Mack, K.-H., Kerp, J., \& Klein, U. 1997 A\&A 324, 870

\noindent
Niemiec, J. \& Ostrowski, M. 2004 ApJ (in press) 

\noindent
Sambruna, R.M., Gliozzi, M., Donato, D., Tavecchio, F., Cheung, C.C., \& Mushotzky, R.F. 2004, A\&A, 414, 885

\noindent
Siemiginowska, A., Bechtold, J., Aldcroft, T.L., Elvis, M.,
Harris, D.E., and Dobrzycki, A.
2002 ApJ 570, 543

\noindent
Spergel, D. N. et al. 2003 ApJS 148, 175

\noindent
Stawarz, L., Sikora, M., Ostrowski, M., \& Begelman, M. C. 2004 ApJ
(in press)\\ http://arXiv.org/abs/astro-ph/0401356

\noindent
Tavecchio, F., Maraschi, L., Sambruna, R.M., \& Urry, C.M. 2000 ApJL
544, L23-26

\noindent
Walker, R.C., Benson, J.M., and Unwin, S.C. 1987 ApJ 316, 546

\noindent
Walker, R.C. 1997 ApJ 488, 675

\noindent
Walker, R.C., Benson, J.M., Unwin, S.C., Lystrup, M.B., Hunter, T.R.,
Pilbratt, G., \& Hardee, P.E. 2001 ApJ 556, 756


\clearpage 

\onecolumn

\begin{figure}
\caption{The HST image used for photometry, with radio contours
 overlaid.  Knot k4 is towards the right hand edge in this figure.
 The bright emission along the southern edge of the jet is part of the
 diffraction spike from the nucleus.  The radio contours are the same
 as used in fig.~\ref{fig:regions}.  The WFPC2 image was taken with the
 F675W filter.
 \label{fig:k4hst}}
\end{figure}

\begin{figure}
\caption{Radio contours at 1.6 GHz.  This map from 1982 data (Walker
  et al. 1987) shows the section of the jet of interest, with several
  knots labelled.  There are additional radio features on larger
  scales (see Walker et al. 1987).  The beamwidth is 3.5$''$ FWHM; the
  first contour level is 0.25 mJy/beam and successive contours
  increase by factors of 2.\label{fig:lband}}
\end{figure}

\begin{figure}
\caption{The 0.5 to 10 keV map of 3C120 with overlayed regions for
measuring the flux.  The X-ray data have been smoothed with a Gaussian
of FWHM= 0.5$^{\prime\prime}$.  Dashed regions are background regions.
Radio contours from the VLA at 5 GHz (CLEAN beamwidth = 0.365") are
overlayed.  Contour levels start at 0.05 mJy/beam and increase by factors
of two.  The regions are labelled by their distance from the core in
arcsec for knots, and by 's2' and 's3' for smooth sections of the jet
detected in the optical by Hjorth et al. (1995).\label{fig:regions}}
\end{figure}

\begin{figure}
\caption{A residual X-ray map after subtraction of the PSF showing the
  excess emission to the southwest of the nucleus.  The white
contours are linear: 0.5, 1.0, 1.5, ... 4.0 counts per
0.049$^{\prime\prime}$ pixel.  An additional contour at -8 shows the
location of the PSF subtraction.  The black area in the center is
negative, most likely caused by pileup suppressing the true peak
intensity.  The image was generated with an energy filter from 0.5 to
10 keV and smoothed with a Gaussian of FWHM=0.25$^{\prime\prime}$.
Overlaid are light magenta radio contours from a 6cm VLA map with the
lowest contour level at 0.2 mJy/beam and successive contours increase
by factors of two.  The beam is circular with
FWHM=0.365$^{\prime\prime}$. \label{fig:inner}}
\end{figure}

\begin{figure}
\plotone{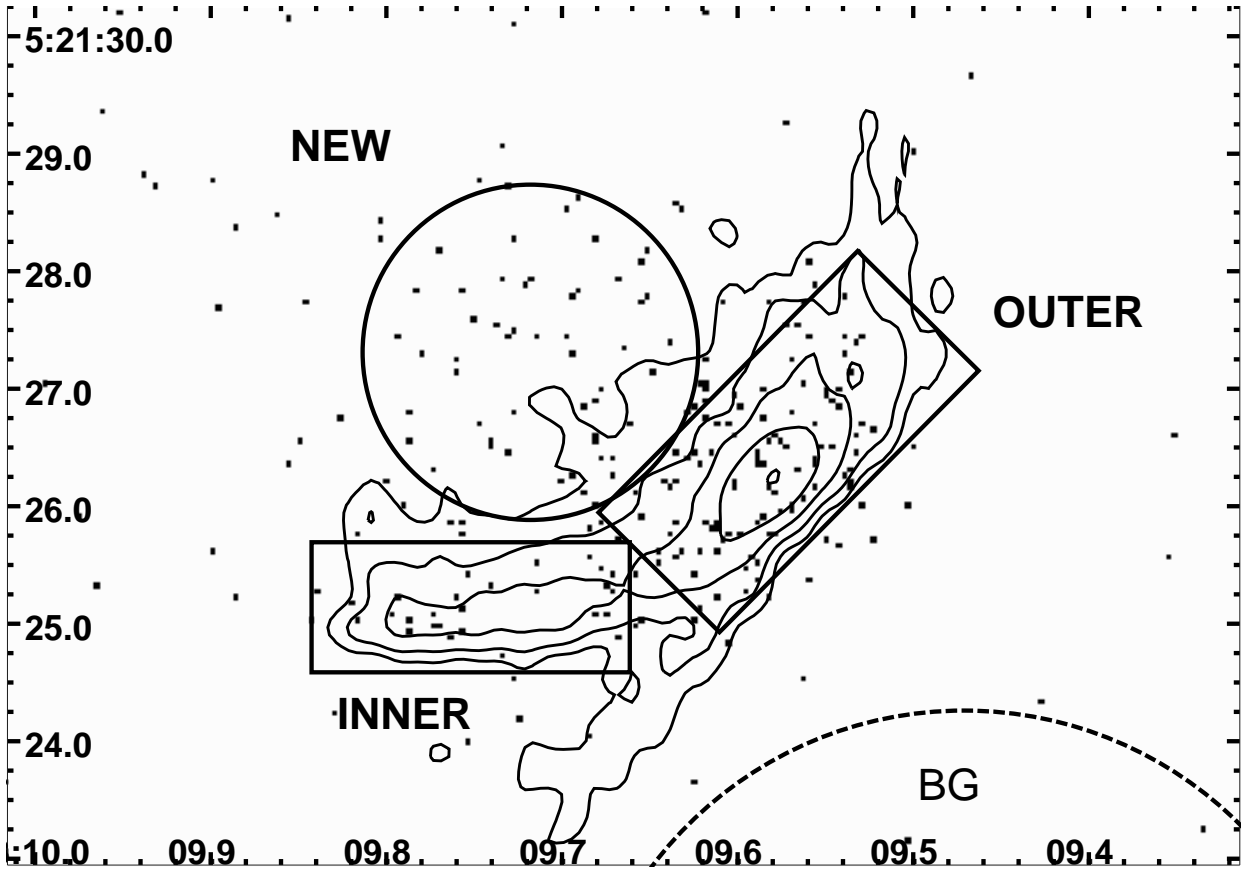}
\caption{The events in the 0.5-10 keV band for k25.  The radio
  contours are the same as used in fig.~\ref{fig:regions}.  The
  regions are shown as well.\label{fig:k25marbles}}
\end{figure}

\begin{figure}
\caption{The outer knot, k80.  The X-ray images come from obsid 1613,
  consist of counts between 0.4 and 6 keV, and have been regrided to
  1/4 native ACIS pixel.  The left panel has been smoothed with a
  Gaussian of FWHM=1$''$; the right panel with a 2$''$ Gaussian.  Radio
  contours are overlayed.  Those on the left show just the brighter,
  more compact levels from a 5~GHz VLA observation with the B array:
  with a beamsize of 1.36$''\times1.13''$ in PA=44$^{\circ}$, and
  contours at 0.15, 0.20, and 0.25 mJy/beam.  The contours on the
  right panel are from a 1.5~GHz (B array) observation with a 3.8$''$
  beam with the first contour at 0.2mJy/beam and successive contours
  increase by factors of two.  There are between 6 and 9 counts in the
  two weak X-ray features upstream of k80.  In both panels, the colors
  span a factor of 4 in X-ray brightness.  The peak brightnesses are
  0.038 (right) and 0.06 (left) counts per 0.123''
  pixel. \label{fig:k80}}
\end{figure}

\begin{figure}
\plottwo{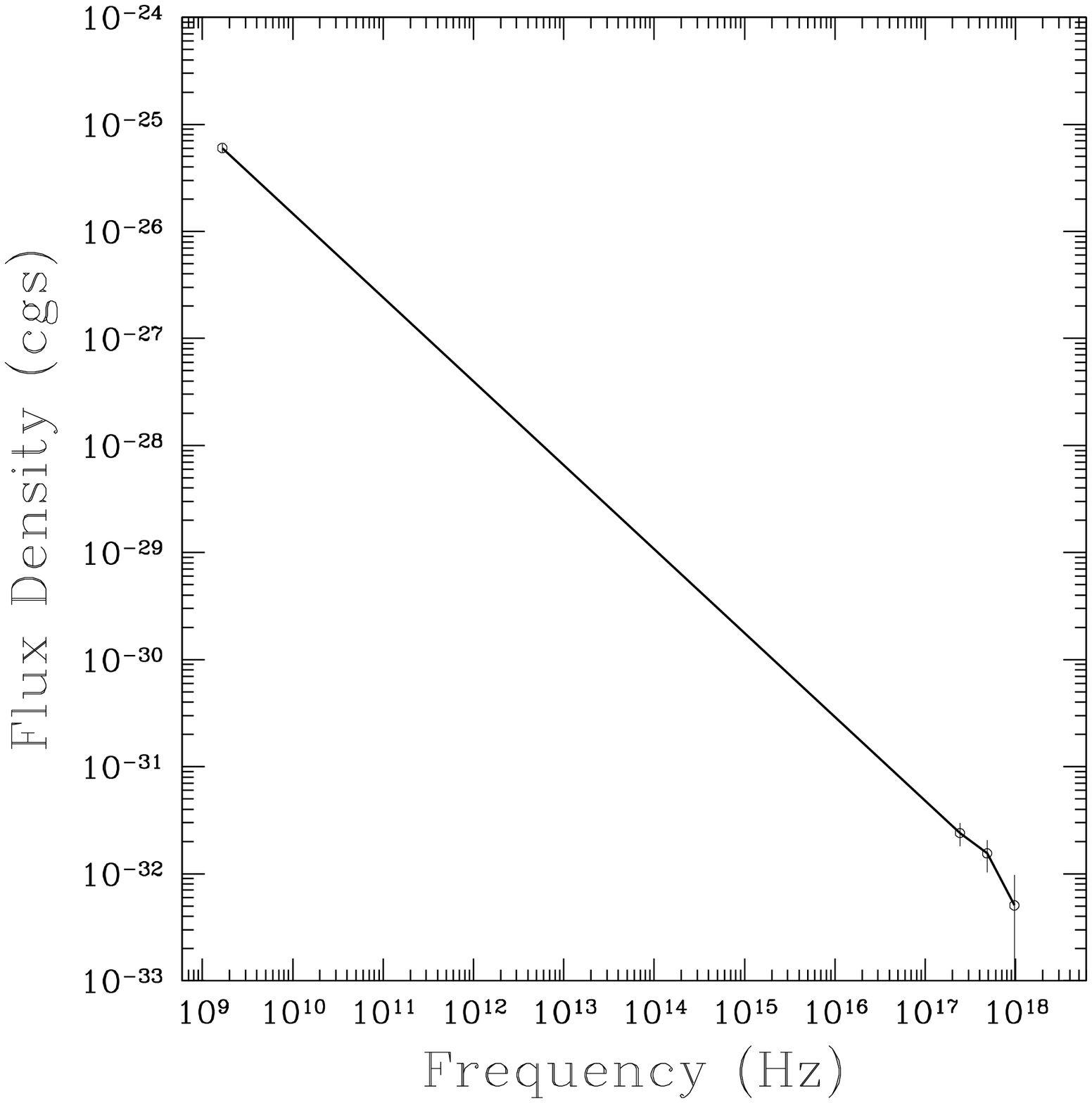}{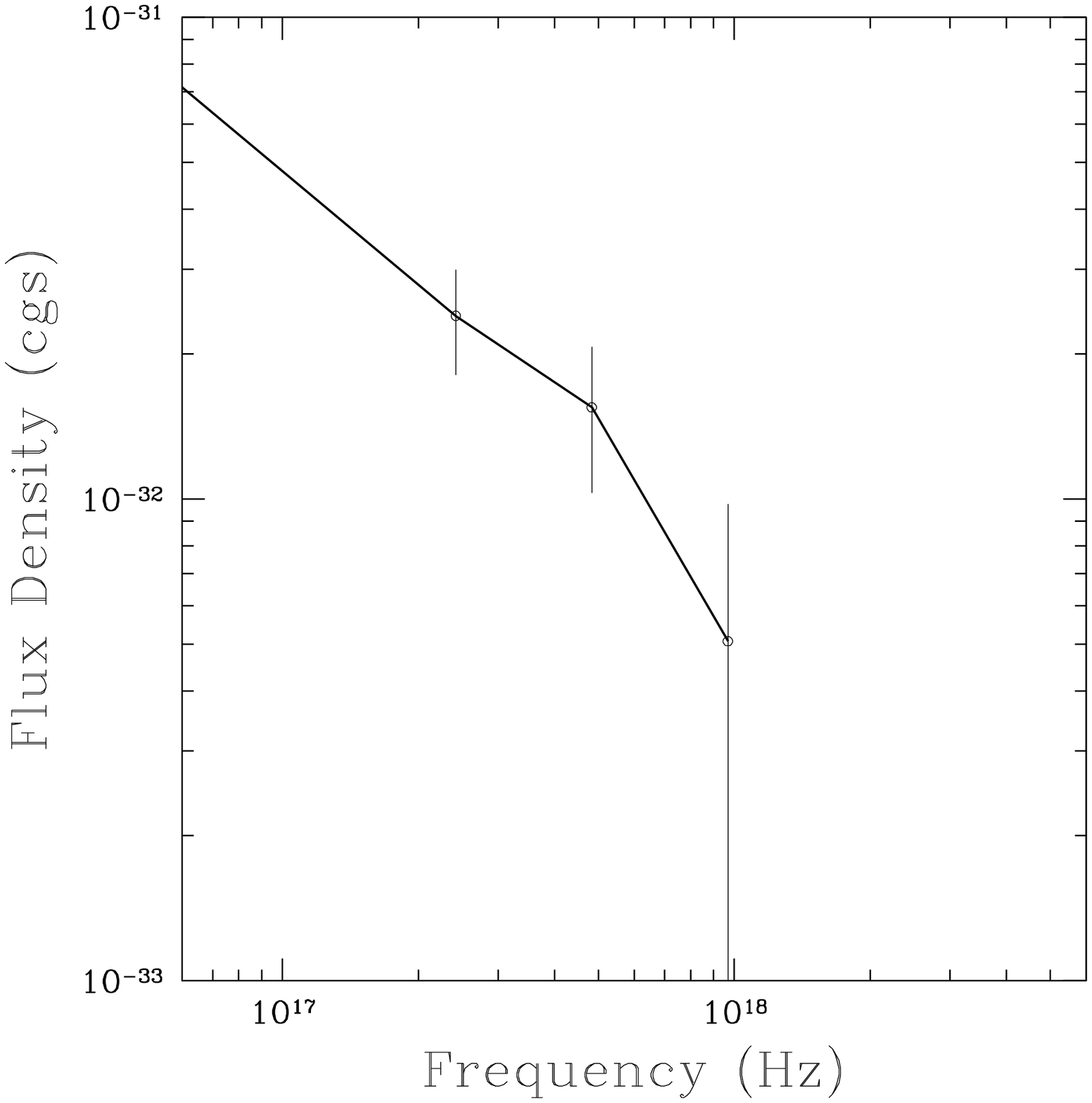}
\caption{The spectrum of k80.  The left panel shows the
  radio to X-ray spectrum; the right panel shows the X-ray
  data. \label{fig:speck80x}}
\end{figure}



\begin{deluxetable}{ccrrrrrrr} 
\tablecolumns{9} 
\tablewidth{0cm} 
\tablecaption{VLA Observations\label{tab:vla}}
\tablehead{
 & & \multicolumn{3}{c}{Beam Size}  & & \multicolumn{2}{c}{Calibrators} \\

\colhead{Frequency} & \colhead{Date(s)} & \colhead{MajAx} &
\colhead{MinAx} & \colhead{PA} & 
\colhead{Peak}  &  \colhead{rms}  & 
\colhead{3C~286}  &\colhead{3C~48} \\

\colhead{(GHz)}     &  \colhead{}        & \colhead{(\arcsec)} & 
\colhead{(\arcsec)} & \colhead{(Deg)} & \colhead{(Jy/beam)}  & \colhead{(mJy)} &
\colhead{(Jy)}  &\colhead{(Jy)} \\

\colhead{(1)} &     \colhead{(2)} & \colhead{(3)} & \colhead{(4)}
& \colhead{(5)} & \colhead{(6)} & \colhead{(7)} & \colhead{(8)} & \colhead{(9)} 
}
\startdata
 1.45  & 1983 Oct. 2 & 1.26 & 1.25 & 47 & 2.76 & 1.45 & 14.51 & \nodata \\
 4.86 & Multiple\tablenotemark{a} & 0.365 & 0.365 & \nodata & 3.15 & 0.012 & 7.55 & \nodata \\
14.94 & 2000 Jan. 4 & 0.54 & 0.51 & 24 & 3.36 & 0.033 & 3.43 & 1.80 \\
43.34 & Multiple\tablenotemark{b} & 2.21 & 1.60 & -2 & 2.07 & 0.25 & 1.45 & 0.575 \\
\enddata
\tablenotetext{a}{Weighted average of images from 1987 Jul. 5, 1991 Aug. 29, and 1996 Nov. 2}
\tablenotetext{b}{Image made from combined uv data from 1999 Apr. 17 and 1999 May 30}

\tablecomments{ The columns in the table give (1) the observing
frequency, (2) the date(s) of the observations, the CLEAN restoring beam
(3) major and (4) minor axis size (FWHM) and (5) position angle, (6)
the peak flux density in the image (always on the 3C~120 core source
which is highly variable), (7) the image off-source rms noise level,
(8) the assumed flux density of 3C~286 used in the flux density
calibration, and (9) the flux density of 3C~48 used in the
calibration, when it was used.}

\end{deluxetable}


\begin{deluxetable}{lllllllll}
\rotate
\tabletypesize{\scriptsize}
\tablecolumns{9} 
\tablewidth{0pc} 
\tablecaption{Flux Densities (ergs~cm$^{-2}$~s$^{-1}$~Hz$^{-1}$) and
  Spectral Indices\label{tab:fluxden}}
\tablehead{ 
\colhead{Quantity} & \colhead{k4} & \colhead{k7} & \colhead{s2} & \colhead{s3} & \colhead{k25in} & \colhead{k25out} & \colhead{k25new} & \colhead{k80}}
\startdata 

S(1.5GHz) & 1.4$\pm$0.1E-24 & 3.2$\pm$0.5E-25 & 1.5$\pm$0.4E-25 & 9.0$\pm$3.9E-26 & 1.2$\pm$0.4E-25 & 1.5$\pm$0.4E-25 & \nodata & 6.0$\pm$0.6E-26\\

S(5GHz) & 6.6$\pm$0.3E-25 & 1.6$\pm$0.1E-25 & 8.4$\pm$0.4E-26 & 5.1$\pm$0.3E-26 & 4.3$\pm$0.2E-26 & 6.9$\pm$0.4E-26 & 9.1$\pm$1.5E-27 & \nodata \\

S(15GHz)	 & 2.8$\pm$0.2E-25 & 7.2$\pm$0.5E-26 & 3.9$\pm$0.3E-26 & 2.6$\pm$0.2E-26 & 2.5$\pm$0.2E-26 & 3.5$\pm$0.3E-26  & \nodata & \nodata\\

S(43GHz)	 & 1.2$\pm$0.1E-25 & 3.2$\pm$0.5E-26 & 1.6$\pm$0.5E-26 & 8.1$\pm$5.0E-27 & 0.9$\pm$0.5E-26 & 1.6$\pm$0.5E-26  & \nodata & \nodata\\

S(4.47E14Hz)	 & 2.7$\pm$0.3E-29 & 0.9$\pm$0.4E-29 & 2.6$\pm$0.4E-29 & 1.0$\pm$0.1E-29  & \nodata & \nodata & \nodata & \nodata\\

S(2.42E17Hz) & 1.0$\pm$0.2E-31 & 6.3$\pm$1.6E-32 & \nodata & 0.8$\pm$0.6E-32 & 1.7$\pm$0.9E-32 & 6.0$\pm$1.5E-32 & 5.5$\pm$1.4E-32 & 2.4$\pm$0.6E-32\\

S(6.05E17Hz) & 5.8$\pm$1.0E-32 & 0.5$\pm$0.4E-32 & 0.5$\pm$0.3E-32 & \nodata & 1.3$\pm$0.4E-32 & 4.7$\pm$0.8E-32 & 2.5$\pm$0.6E-32 & 1.6$\pm$0.5E-32\tablenotemark{a}\\

S(9.67E17Hz)	 & 3.8$\pm$0.8E-32 & 0.2$\pm$0.2E-32 & \nodata & \nodata & 1.4$\pm$0.5E-32 & 5.8$\pm$0.9E-32 & 2.4$\pm$0.5E-32 & 5.1$\pm$4.7E-33\\

S(1.69E18Hz)	 & 2.7$\pm$1.1E-32 & \nodata & 0.2$\pm$0.1E-32 & \nodata & 1.5$\pm$0.7E-32 & 3.5$\pm$1.2E-32 & 1.0$\pm$0.6E-32 & \nodata \\
\cutinhead{Spectral Indexes}
$\alpha_r$ & 0.74$\pm$0.04 & 0.68$\pm$0.04 & 0.67$\pm$0.06 & 0.69$\pm$0.10 & 0.74$\pm$0.06 & 0.66$\pm$0.03 & \nodata & \nodata \\

$\alpha_{rx}$ & 0.90 & 0.86 & 0.97 & \nodata & 0.82 & 0.81 & 0.70 & 0.79 \\

$\alpha_x$ & 0.90$\pm$0.20\tablenotemark{b} & 2.4$\pm$0.6 & 0.2$\pm$0.6 & \nodata & 0.0$\pm$0.4 & 0.3$\pm$0.3 & 0.6$\pm$0.2 & 0.7$\pm$0.4 \\
\enddata


\tablenotetext{a}{This flux density refers to a frequency of 4.8E17~Hz
(2 keV).}

\tablenotetext{b}{The value in the table comes from the 4 band flux densities with the soft bands corrected for excess absorption (Nh=2.4$\times10^{21}~cm^{-2}$).  The galactic column density in this direction is 1.08$\times10^{21}~cm^{-2}$.  Sherpa spectral fitting
gives $\alpha_x$=0.95$\pm$0.45 (Nh=(1.9$\pm$2.2)$\times10^{21}~cm^{-2}$) for no background subtraction and 1.01$\pm$0.6 (Nh=(2.4$\pm$3.1)$\times10^{21}~cm^{-2}$) with background subtraction.}

\end{deluxetable}


\begin{table}
\begin{center}
\caption{Chandra observations\label{tab:xlog}}
\bigskip
\begin{tabular}{llllll}
\tableline\tableline
Obsid  &     Date  &  	Detector  &  	Grating  &  	livetime  &     CALDB \\
       &           &              &              &     (sec)  &    \\
\tableline
1613   &   2001-09-18  &  ACIS-235678  &  none  &  12,881  &  	2.8 \\
3015    &  2001-12-21  &  ACIS-456789  &  HETG  &  57,218  &  	2.10 \\
\tableline
\end{tabular}



\tablecomments{Both observations were made with the standard CCD
readout time of 3.2s.}
\end{center}
\end{table}


\begin{table}
\begin{center}
\caption{Specifics of the regions and assumed sizes for volumes\label{tab:regions}}
\begin{tabular}{llllllll}
\tableline\tableline
 & \multicolumn{4}{c}{Region Specification} & \multicolumn{3}{c}{Assumed Volumes}\\
region  &  shape  & RA(J2000)  & DEC(J2000)  &   size\tablenotemark{a}  &   shape  &
dimensions\tablenotemark{b}  &   pathlength\\
  &    &    &    &  ($^{\prime\prime}$)  &    &  ($^{\prime\prime}$)
&  (kpc) \\
\tableline
k4  &  	circle  &  04:33:10.843  &  +05:21:15.09  & 0.738  &  sphere  &  0.33  &  0.42 \\
k7  & 	rotbox	  & 04:33:10.628  & +05:21:15.44  & 2.165, 1.427  & sphere  & 0.33  & 0.42\\
s2  & 	rotbox	  & 04:33:10.450  & +05:21:16.70  & 3.542, 1.230  & cyl.  & 0.33;3.0  & 0.42\\
s3  & 	rotbox   &  04:33:10.232  & +05:21:18.35  & 3.444, 1.230  & cyl.  & 0.33;2.8  & 0.42\\
k25in  & rotbox	  & 04:33:09.752  & +05:21:25.14  & 2.706, 1.107  & cyl.  & 0.33;2.1  & 0.42\\
k25out  & rotbox  & 04:33:09.571  & +05:21:26.55  & 3.132, 1.447  & cyl.  & 0.6;1.8  & 0.76\\
k25new  & circle  & 04:33:09.718  & +05:21:27.31  & 1.427  & sphere  & 1.43  & 1.82\\
k80  & 	circle	  & 04:33:07.937  & +05:22:21.22  & 4.797  & sphere  & 3.0  & 3.82\\

\tableline
\end{tabular}

\tablenotetext{a}{A single entry denotes radius of a circle; two
entries are sides of the rectangle.}
\tablenotetext{b}{Radius of sphere or cylinder.  When a second
entry is present, it is the length of the cylinder.}

\tablecomments{Columns to the left describe the regions used for
photometry except where noted in the text.  The last 3 columns give
the assumed geometry of the emitting volumes.  The background regions
are shown in fig.~\ref{fig:regions}.  All coordinates refer to the
radio maps.}
\end{center}
\end{table}



\begin{deluxetable}{lrrrrrrrrr}
\rotate
\tabletypesize{\small}
\tablecaption{X-ray fluxes (in units of
10$^{-14}$~erg~cm$^{-2}$~s$^{-1}$) and source counts\label{tab:xflux}}
\tablewidth{0pt}
\tablehead{
\colhead{Energy\tablenotemark{a}}  &  \colhead{k4}  &  \colhead{k7}   &     \colhead{s2}    &    \colhead{s3}   &     \colhead{k25in}  &   \colhead{k25out}  &  \colhead{k25new}  &  \colhead{k25tot} &  \colhead{k80\tablenotemark{b}}
}

\startdata

0.5-1.5 &  1.85$\pm$0.39 &  1.16$\pm$0.27 &  \nodata &  0.15 &     0.31$\pm$0.16 & 
1.11$\pm$0.28 &  1.02$\pm$0.25 &  3.15$\pm$0.47 & 0.49$\pm$0.12\\

1.5-3 & 3.35$\pm$0.57 &  0.26$\pm$0.22 &  0.27$\pm$0.16 & \nodata  &    0.77$\pm$0.25 & 
2.72$\pm$0.49 &  1.43$\pm$0.36 &  6.03$\pm$0.72 & 0.50$\pm$0.17\\

3-6 & 2.55$\pm$0.56 &  0.15$\pm$0.12 &  \nodata   &   0.14 &     0.93$\pm$0.31 & 
3.90$\pm$0.62 &  1.60$\pm$0.37 &  7.80$\pm$0.86 & 0.34$\pm$0.32\\

6-10 & 2.35$\pm$0.96 &  \nodata   &  0.16$\pm$0.10 &  \nodata  &    1.29$\pm$0.65 & 
3.02$\pm$1.03 &  0.90$\pm$0.54 &  6.10$\pm$1.40 & ...\\

0.5-10  & 10.1$\pm$0.91 &   1.42$\pm$0.36 &  0.43$\pm$0.26 &  \nodata &     3.30$\pm$0.99 & 
10.7$\pm$1.07 &  4.95$\pm$0.64 &  23.08$\pm$1.62 & 1.33$\pm$0.39\\

\cutinhead{Source Counts}

Total counts & 124  &     34   &     14   &     7   &      27   &     98
 &  55   &   225 & 48\\
Net counts &  121   &    29   &      7   &     0   &      22   &     90 &
 44 &  215 & 37\\
\enddata

\tablenotetext{a}{The energy band is given in keV.}
\tablenotetext{b}{The measurements for k80 come from obsid 1615 and
the 3 bands are 0.4-1.5keV, 1.5-3keV, and 3-6 keV; with the broad band
being the sum of these 3.}

\end{deluxetable}


\normalsize





\begin{deluxetable}{lrlllrrrrr} 
\tablecolumns{10} 
\tablewidth{0pc} 
\tablecaption{Parameters for a Synchrotron Model\label{tab:sync}}
\tablehead{ 
\colhead{} & \colhead{} &
\colhead{} & \colhead{} & \colhead{} & \colhead{} &
\multicolumn{2}{c}{$\delta=$3\tablenotemark{a}} &
\multicolumn{2}{c}{$\delta=$6\tablenotemark{a}} \\ 
\cline{7-8}
\cline{9-10}\\

\colhead{Region} & \colhead{ B(1)\tablenotemark{b}} &
\colhead{$\alpha_r$\tablenotemark{c}} &
\colhead{$\alpha_{rx}$\tablenotemark{d}} & \colhead{logL$_s$} &
\colhead{$\tau_{\frac{1}{2}}$\tablenotemark{e}} & \colhead{logL$_s$} &
\colhead{$\tau_{\frac{1}{2}}$} & \colhead{logL$_s$} &
\colhead{$\tau_{\frac{1}{2}}$}\\ 

\colhead{} & \colhead{($\mu$G)} &
\colhead{} & \colhead{} & \colhead{(erg/s)} & \colhead{(years)} &
\colhead{(erg/s)} & \colhead{(years)} & \colhead{(erg/s)} &
\colhead{(years)}} 

\startdata 

k4 & 116 & 0.74 & 0.9 & 41.83 & 23 & 39.92 & 56 & 38.72 & 27\\

k7 & 72 & 0.68 & 0.84 & 41.47 & 47 & 39.56 & 86 & 38.36 & 47\\

s2 & 33 & 0.67 & 0.97 & 40.62 & 152 & 38.73 & 108 & 37.53 & 17\\

k25inner& 36 & 0.74 & 0.82 & 41.14 & 131 & 39.23 & 108 & 38.03 &
18\\

k25outer& 27 & 0.66 & 0.81 & 41.45 & 195 & 39.54 & 105 & 38.34 &
16\\

k25new & 9 & \nodata & 0.70 & 41.24 & 669 & 39.33 & 74 & 38.13 &
9\\

k80 & 8.3 & \nodata & 0.79 & 41.14 & 762 & 39.24 & 69 & 38.03 & 9\\
\enddata

\tablenotetext{a}{$\delta$ is the beaming factor}

\tablenotetext{b}{B(1) is the equipartition magnetic field strength for the case
$\delta=$1.  For other values of $\delta$, it is B(1)/$\delta$ (Harris
\& Krawczynski, 2002).}

\tablenotetext{c}{$\alpha_r$ is the radio spectral index.}

\tablenotetext{d}{$\alpha_{rx}$ is the spectral index connecting the
radio to the X-ray bands.}

\tablenotetext{e}{$\tau_{\frac{1}{2}}$ is the halflife, as observed at the earth, for
the electrons responsible for the observed 2keV emission (eq.B7 of 
Harris \& Krawczynski, 2002, assuming $\Gamma=\delta$).}


\tablecomments{The emission spectrum is assumed to cover the range E6 to 5E18 Hz
in the observed frame.}

\end{deluxetable}



\begin{table}
\begin{center}
\caption{Parameters for an Inverse Compton Model with Bulk
  Relativistic Velocities\label{tab:ic}}
\begin{tabular}{llrlrrrr}
\tableline\tableline
Region & $\alpha$\tablenotemark{a} &  B(1)\tablenotemark{b} & R(1)\tablenotemark{c} & $\delta$\tablenotemark{d} & $\theta$ & R$'$\tablenotemark{e} & Distance\tablenotemark{f}\\
 &  & $\mu$G &  & $\Gamma$ & (deg) & 	 & (kpc)\\
\tableline
k4      & 0.74  & 116   & 0.0917 & 16   & 3.5   & 65    & 42\\
k7	& 0.68	&  72	& 0.0176 &  12	&   4.5	&   56	&  57 \\
s2	& 0.67	&  33	& 0.0104 & 7.7  & 7.5   & 40    & 49 \\
k25in	& 0.74	&  36	& 0.3967 & 13	&  4.5	&  352	& 184\\
k25out	& 0.66	&  27	& 0.1939 & 18	&  3.2	& 1840	& 269\\
k25new	& 0.70	&  10	& 1.3454 & 14	&  4.1	& 5060	& 215\\
k80	& 0.79	&   8.3	& 0.6289 &  5.3	& 10.5	&  137	& 281\\
\tableline
\end{tabular}


\tablenotetext{a}{$\alpha$ is the spectral index for both the IC and
  synchrotron spectra.  For most features, it is $\alpha_r$ from
  Table~\ref{tab:fluxden}.  For k25new and k80, it is $\alpha_{rx}$.}

\tablenotetext{b}{B(1) is the equipartition magnetic field strength for no beaming.}

\tablenotetext{c}{R(1) is the ratio of amplitudes of the IC to
synchrotron (observed) spectra.}

\tablenotetext{d}{The $\delta$/$\Gamma$ column gives their values for
$\Gamma$=$\delta$, and the relevant angle to the line of sight is
$\theta$.}

\tablenotetext{e}{R$'$ is the ratio of E$^2$ losses in the jet frame
  (IC and synchrotron losses).}

\tablenotetext{f}{'Distance' is the de-projected distance from the
core for the particular feature (projected distance/sin$\theta$).}


\tablecomments{For k7, the X-ray spectrum is much steeper than the radio spectrum
so the IC/CMB model is not applicable.  For both k25in and k25out, the
actual X-ray spectrum is significantly flatter than the radio
spectrum, so also for these features, the IC/CMB beaming model is not
tenable.}
\end{center}
\end{table}




\begin{table}
\begin{center}
\caption{Thermal Bremsstrahlung Parameters\label{tab:thermal}}
\begin{tabular}{lllllr}
\tableline\tableline
Region & n$_e$\tablenotemark{a} & Cooling Time & Mass  & Pressure\tablenotemark{b} & RM\tablenotemark{c}\\
      & (cm$^{-3}$) & (years) & (M$_{\odot}$) & (dyne~cm$^{-2}$)  & (m$^{-2}$)\\
\tableline
k4 & 6.6 & 1.5E5 & 6.3E6 & 2.0E-8 & 22,450\\
k7 & 2.5 & 4.0E5 & 2.4E6 & 7.5E-9 &  8,500\\
s2 & 0.6 & 1.8E6 & 3.7E6 & 1.7E-9 & 2,040\\
k25inner & 1.7 & 5.6E5 & 8.0E6 & 5.6E-9 & 5,780\\
k25outer & 1.9 & 5.2E5 & 2.4E7 & 5.6E-9 & 11,450\\
k25new	 & 0.5 & 1.9E6 & 4.0E7 & 1.5E-9 & 7,520\\
k80      & 0.08 & 1.2E7 & 5.9E7 & 2.5E-10 & 2,500\\
\tableline
\end{tabular}


\tablenotetext{a}{The electron density is that necessary for a uniform plasma
with the volume described in Table~\ref{tab:regions}.}

\tablenotetext{b}{The pressure values assume a temperature of 1 keV.}

\tablenotetext{c}{RM gives the predicted rotation measure for a field
of 10 $\mu$G and a pathlength from table~\ref{tab:regions}.
RM(radians~m$^{-2}$)=810$\times$B$(\mu$G)$\times$dL(kpc)$\times$n$_e$(cm$^{-3}$).}

\end{center}
\end{table}

\end{document}